\documentclass[fleqn,12pt,twoside]{article}
\usepackage[headings]{espcrc1}

\readRCS
$Id: espcrc1.tex,v 1.2 2004/02/24 11:22:11 spepping Exp $
\usepackage{graphicx,wrapfig}
\usepackage[figuresright]{rotating}

\newcommand{\AmS}{{\protect\the\textfont2
  A\kern-.1667em\lower.5ex\hbox{M}\kern-.125emS}}

\title{Critical end point and its consequences}

\author{Masayuki Asakawa\address{Department of Physics, Osaka University,
        Toyonaka 560-0043, Japan}%
        \thanks{This work was in part supported by
           Grant-in-Aid by the Japanese Ministry of Education
           Nos. 14540255 and 17540255.}
        and
        Chiho Nonaka\address{Department of Physics, Duke University,
                             Durham, NC 27708, U.S.A.}
        \thanks{Present address: Department of Physics and Astronomy, 
	  116 Church Street S.E., University of Minnesota, 
	  Minneapolis, MN 55455, U.S.A.}
        \thanks{This work was in part supported by DOE grants
          DE-FG02-96ER40945 and DE-FG02-03ER41239.}}

\runtitle{Critical end point and its consequences}
\runauthor{M. Asakawa and C. Nonaka}
\begin{document}

\maketitle

\setcounter{footnote}{0}

\begin{abstract}
Recently a lot of evidence that there exists
a critical end point (CEP) in the QCD phase diagram has been accumulating.
However, so far, no reliable equation of state with the CEP has been
employed in hydrodynamical calculations. In this article, we construct
the equations of state with the CEP on the basis of the universality
hypothesis and show that the CEP acts as an attractor
of isentropic trajectories. We also consider the time evolution
in the case with the CEP and discuss how the CEP affects the final
state observables. 
\end{abstract}

\section{UNIVERSALITY AND EQUATION OF STATE}
The existence of the CEP in the
QCD phase diagram has been predicted by several effective theory
analyses \cite{AsaYa89,St}. Recent lattice calculation has
also shown it \cite{FoKa02}, although the lattice size is still rather small
and recently a question on the procedure in Reference \cite{FoKa02}
was raised \cite{ejiri05}.
In this article, we construct equations of state with the CEP
on the basis of the universality, which is a general principle
in phase transition theory and discuss the consequences of the CEP
in final state observables in high energy heavy ion collisions.
See Reference \cite{NonakaAsakawa} for further
details of the formulation and calculation.

The equation of state with the CEP consists of two parts,
the singular part and non-singular part. We assume that the CEP
in QCD belongs to the same universality class as that in
the three dimensional Ising model on the basis of the
universality hypothesis. After mapping the variables and
the equation of state near
the CEP in the three dimensional Ising model onto those in QCD,
we match the singular entropy density near the
CEP with the non-singular QGP and hadron phase
entropy densities which are known away from the CEP.
From this procedure we determine the behavior of the entropy density
which includes both the singular part and non-singular part
in a large region in the $T$-$\mu_B$ plane.

In the three dimensional Ising model, the magnetization $M$
(the order parameter)
is a function of the reduced temperature $r=(T-T_c)/T_c$ and
the external magnetic field $h$ with $T_c$ being
the critical temperature.
The CEP is located at the origin
$(r,h)=(0,0)$.
At $r<0$ the order of the
phase transition is first and at $r>0$
it is crossover.

In order to determine the singular part of the entropy density,
we start from the Gibbs free energy density $G(h,r)$,
\begin{equation}
G(h,r) = F(M,r)-Mh,
\end{equation}
where $F(M,r)$ is the free energy density.
Differentiating the Gibbs free energy by the temperature,
we obtain the singular part of the entropy density $s_c$,
\begin{equation}
s_c  =  - \left ( \frac{\partial G}{\partial T} \right )_{\mu_B}.
\label{Eq-Sc}
\end{equation}
Note that $T$ in Eq. (\ref{Eq-Sc}) is the temperature on
the QCD side.

Next, the mapping between the $r$-$h$ plane in the three dimensional Ising model
and the $T$-$\mu_B$ plane in QCD needs to be specified.
The CEP in the three dimensional Ising Model,
which is the origin in the $r$-$h$ plane, is mapped to the CEP in QCD,
$(T,\mu_B) = (T_E,\mu_{BE})$.
The $r$ axis is tangential to the first order phase transition
line at the CEP \cite{GeRt}.
There is no general rule about how the $h$ axis is
mapped in the $T$-$\mu_B$ plane.
For simplicity, we set the $h$ axis perpendicular to the $r$ axis.
For quantitative construction of equations of state
with the CEP,
we fix the relation between the scales in
$(r,h)$ and $(T,\mu_B)$ variables,
which provides the size of the critical region
around the CEP in the $T$-$\mu_B$ plane, as follows:
$\Delta r = 1$ ($r$-$h$ plane) $\leftrightarrow \Delta \mu_{B\rm crit}$
($T$-$\mu_B$ plane) and
$\Delta h = 1$ ($r$-$h$ plane) $\leftrightarrow
\Delta T_{\rm crit}$ ($T$-$\mu_B$ plane).
In order to connect the equations of state in the singular
region and the non-singular region smoothly,
we define the dimensionless variable $S_c(T, \mu_B)$ for the singular
part of the entropy density $s_c$, which has the dimension [energy]$^{-1}$,
\begin{equation}
S_c(T,\mu_B) = A(\Delta T_{\rm crit}, \Delta \mu_{B\rm crit})
s_c(T,\mu_B),
\label{Eq-size}
\end{equation}
where $A(\Delta T_{\rm crit}, \Delta \mu_{B\rm crit})
=\sqrt{\Delta T^2_{\rm crit}  + \Delta \mu^2_{B\rm crit}} \times D$
and $D$ is a dimensionless constant.
The extension of the critical domain around the CEP is specified
by the parameters $\Delta T_{\rm crit}$,
$\Delta \mu_{B\rm crit}$, and $D$.

Using the dimensionless variable
$S_c(T, \mu_B)$,
we define the entropy density in the $T$-$\mu_B$ plane,
\begin{equation}
s (T, \mu_B)  =  \frac{1}{2}
\left( 1 - \tanh [S_c (T, \mu_B) ]\right ) s_{\rm H}(T, \mu_B)
 + \frac{1}{2}
\left( 1 + \tanh[S_c (T, \mu_B)] \right ) s_{\rm Q}(T, \mu_B),
\label{Eq-rentro}
\end{equation}
where $s_{\rm H}$ and $s_{\rm Q}$ are the entropy densities in the
hadron phase and QGP phase away from the CEP, respectively.
This entropy density includes both singular and non-singular
contributions, and more importantly, gives the correct
critical exponents near the QCD critical end point.
All thermodynamical quantities are obtained from the
entropy density. In the following, we use
$s_{\rm H}$ calculated from the equation of state
of the hadron phase in the excluded volume approximation \cite{RiGo}
and $s_{\rm Q}$ obtained from 
the equation of state of the QGP phase in the Bag model.

\section{RESULTS and DISCUSSIONS}

When entropy production can be
ignored, the entropy and baryon number are conserved in each
volume element and, therefore, the temperature and
chemical potential in a given volume element change
along the contour lines specified by the initial condition.

\begin{figure}[thb]
\begin{minipage}[t]{75mm}
\includegraphics[width=1.0\linewidth]{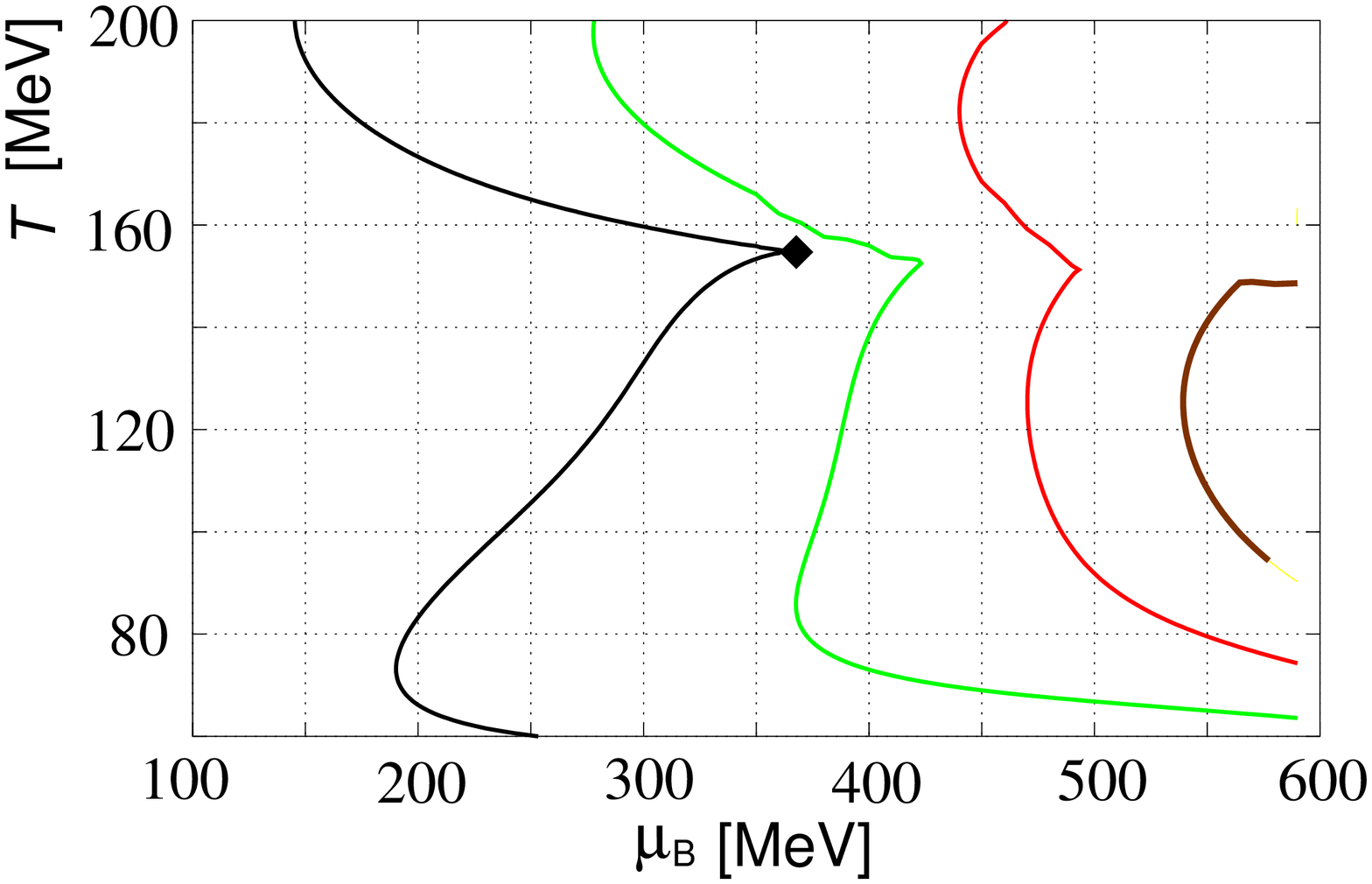}
\caption{Isentropic trajectories in the cases with the CEP.
The CEP is located at $(T_{\rm E}, \mu_{B{\rm E}}) =
(154.7 \hspace{2mm}{\rm MeV},
~367.8 \hspace{2mm}{\rm MeV})$.
The values of $n_B/s$ on the trajectories are
0.01, 0.02, 0.03, and 0.04 from left to right.}
\label{Fig-nbs}
\end{minipage}
\hspace{\fill}
\begin{minipage}[t]{75mm}
\includegraphics[width=1.0\linewidth]{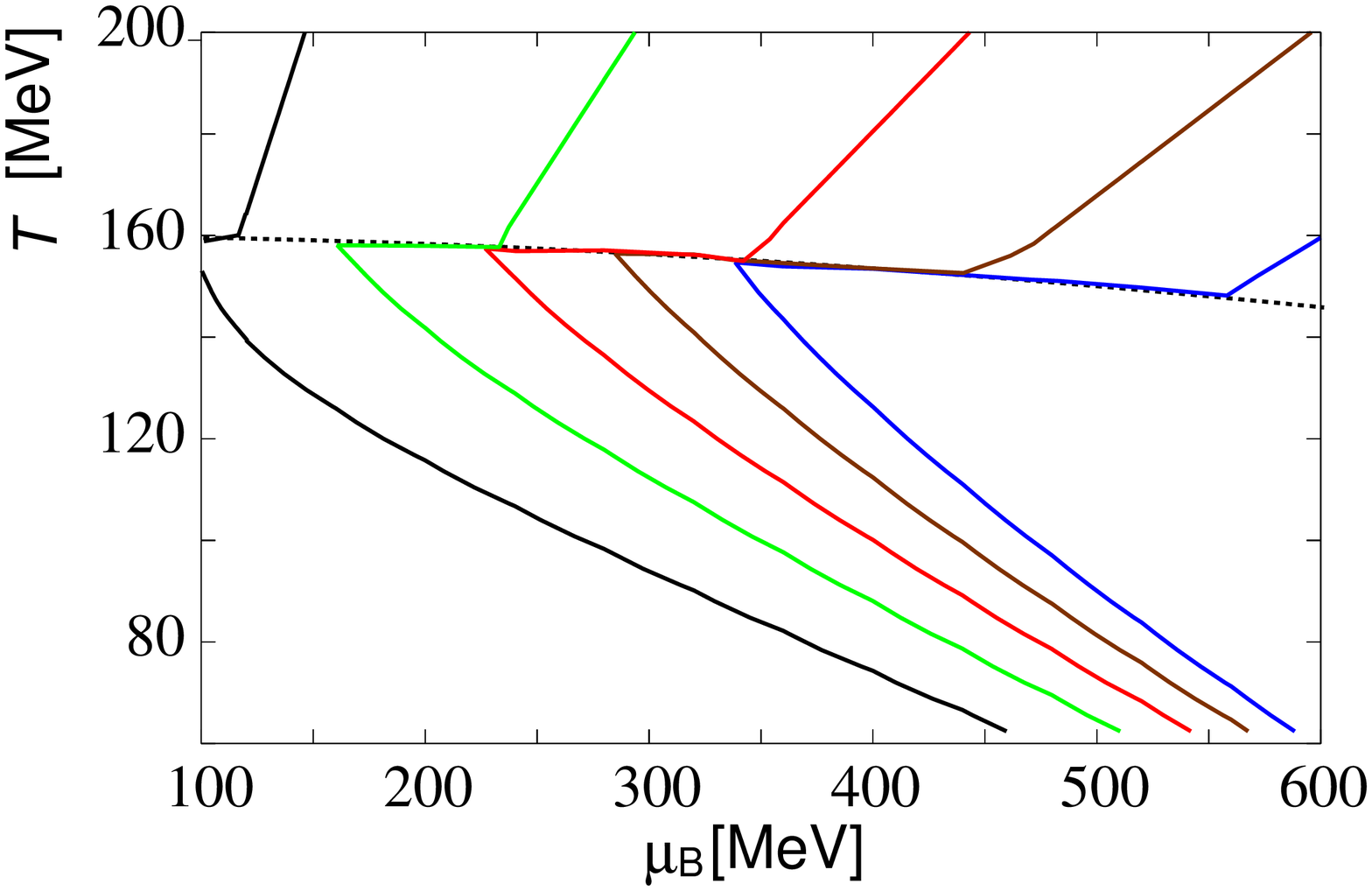}
\caption{Isentropic trajectories (solid lines) in the bag plus excluded volume
model. The values of $n_B/s$ on the trajectories are
0.01, 0.02, 0.03, 0.04, and 0.05 from left to right.
The dashed line stands for the first order phase boundary.}
\label{Fig-nbsc}
\end{minipage}
\vspace*{-0.2cm}
\end{figure}

Figure \ref{Fig-nbs} shows the isentropic trajectories
in the $T$-$\mu_B$ plane, i.e., contour lines of $n_B/s$.
The values of $(\Delta T_{\rm crit},~\Delta \mu_{B{\rm crit}},~D)$ are
$(100~{\rm MeV},~200~{\rm MeV},~0.15)$.
The trajectories are focused
to the CEP. 
Thus, the CEP acts as an attractor of
isentropic trajectories.
Figure \ref{Fig-nbsc} shows isentropic
trajectories in the bag plus excluded volume model,
which is currently employed in most of hydrodynamical
calculations.
There is no focusing effect on
the isentropic trajectories in this case.
Instead, the trajectories are just shifted to the left
on the phase transition line.
Thus, the hydrodynamical
evolution in the case with the CEP
is very different from the one in the case with
the equation of state given by the bag plus excluded volume model.
This attractor character of the CEP leads to the following
consequences:
it is not needed to fine-tune the collision energy
to make the system pass near the CEP and 
the effect of the CEP in observables, if any,
changes only slowly as the collision energy is changed.
We note that the entropy densities
in both the hadron phase and
QGP phase must be carefully taken into account in order to consider
this focusing effect. It is because the baryon number density is 
given by the integral of $\partial s/\partial \mu_B$ with regard to
the temperature. If one fails to reproduce the entropy density
in the hadron phase, it in turn affects the baryon number density
in the QGP phase, and the focusing property of the isentropic trajectories.
See, for example, Reference \cite{BKQM2005}.

\begin{wrapfigure}{l}{75mm}
\begin{flushright}
\vspace*{-0.9cm}
\includegraphics*[width=1.\linewidth]{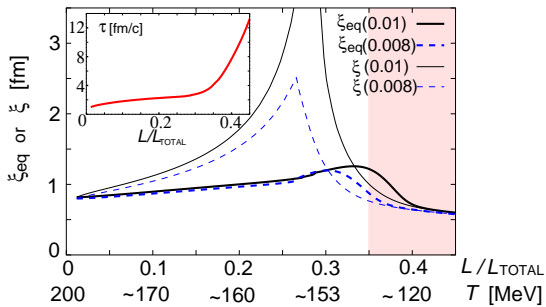}
\caption{
Equilibrium correlation length $\xi_{\rm eq}$ and
non-equilibrium correlation length $\xi$
on the isentropic trajectories
with $n_B/s=0.008$ and $0.01$, together with $\tau$ as functions
of $L/L_{\rm total}$ (inlet).}
\vspace*{-1.1cm}
\label{Fig-cor}
\end{flushright}
\end{wrapfigure}

Finally, we present the time evolution of the correlation length.
Figure \ref{Fig-cor} shows the correlation length as a function of
$L/L_{\rm total}$, where $L$ is the path length to a point
along the isentropic trajectory with a given $n_B/s$ from
a reference point on the same isentropic trajectory on the $T$-$\mu_B$
plane and $L_{\rm total}$ is the one to another reference point along the
trajectory. The dashed and solid lines stand for the correlation lengths
in equilibrium at $n_B/s=0.008$ and $n_B/s=0.01$,
respectively. The thin and thick lines are the equilibrium
correlation length $\xi_{\rm eq}$ and correlation length $\xi$,
respectively. We assume that the thermal equilibrium is
established soon after collisions and that the medium follows
Bjorken's scaling solution. The initial temperature and proper
time are set to 200 MeV and 1 fm/$c$, respectively.
The other parameters are the same as for Figure \ref{Fig-nbs}.
The maximum value of $\xi_{\rm eq}$ along the former
trajectory is larger than that along the latter, because the
former approaches the CEP more closely than the latter.

While the equilibrium correlation length diverges at the CEP,
the actual correlation length does not diverge because of the
critical slowing down. Furthermore, the correlation length at
freezeout does not have to show substantial enhancement even if
the system passes through the CEP before it freezes out. It is
due to the final state interaction in the hadron phase.
Figure \ref{Fig-cor} shows this clearly. 
The non-equilibrium correlation length $\xi$ is smaller than
$\xi_{\rm eq}$ at the beginning.
Then, $\xi$ becomes larger than $\xi_{\rm eq}$ later.
These are both due to the critical slowing down around the CEP,
as pointed out in Reference \cite{BeRa}. However, the difference becomes
small by the time the system gets to the
kinetic freezeout point. If the transverse expansion is taken
into account, the time scale in the hadron phase becomes much shorter,
but $\xi_{\rm eq}$ is already small in the hadron phase and
the difference is expected to remain small.
The dependence on the non-universal constant $A$, the definition
of which is given in References \cite{NonakaAsakawa,BeRa}, is very weak.
Thus, even if there is a CEP in the QCD phase diagram, sudden increase
in the correlation length and fluctuation as a function of collision
energy is not expected. Instead, the low kinetic freezeout temperature
like that observed at RHIC is anticipated on the left hand side of the
CEP. See Reference \cite{NonakaAsakawa} for the details.
Finally, we note that the focusing of the isentropic trajectories
does not necessarily lead to the focusing of the chemical freezeout
points, since the critical region around the CEP, where the phase transition is
of second order, is the region where the free-resonance gas model is
by no means valid.

\end{document}